\documentclass{ecai}
\usepackage{times}
\usepackage{graphicx}
\usepackage{latexsym}
\usepackage{wrapfig}
\usepackage{url}

\graphicspath{{pix/}}

\usepackage[T1]{fontenc}
\usepackage[utf8]{inputenc}

\usepackage{listings}
\usepackage{xcolor}
\usepackage{siunitx}

\usepackage[colorlinks=true,allcolors=black]{hyperref}

\definecolor{codegray}{rgb}{0.4,0.4,0.4}
\definecolor{backcolour}{rgb}{0.98,0.98,0.98}
\definecolor{codepurple}{RGB}{127,0,85}
\definecolor{darkolivegreen}{rgb}{0.33,0.42,0.18}
\colorlet{punct}{red!60!black}
\definecolor{delim}{RGB}{20,105,176}
\colorlet{numb}{magenta!60!black}

\lstdefinelanguage{json}{
    string=[s]{"}{"},
    comment=[l]{:\ "},
    morecomment=[l]{:"},
    morecomment=[l]{\#},
    literate=
        *{0}{{{\color{numb}0}}}{1}
         {1}{{{\color{numb}1}}}{1}
         {2}{{{\color{numb}2}}}{1}
         {3}{{{\color{numb}3}}}{1}
         {4}{{{\color{numb}4}}}{1}
         {5}{{{\color{numb}5}}}{1}
         {6}{{{\color{numb}6}}}{1}
         {7}{{{\color{numb}7}}}{1}
         {8}{{{\color{numb}8}}}{1}
         {9}{{{\color{numb}9}}}{1}
         {Int}{{{\color{numb}Int}}}{3}
         {Bool}{{{\color{numb}Bool}}}{4}
         {Float}{{{\color{numb}Float}}}{5}
         {Double}{{{\color{numb}Double}}}{6}
         {String}{{{\color{numb}String}}}{6}
         {Object}{{{\color{numb}Object}}}{6}
         {Event}{{{\color{numb}Event}}}{5}
         {Detail1}{{{\color{numb}Detail1}}}{7}
         {Detail2}{{{\color{numb}Detail2}}}{7}
         {Value1}{{{\color{numb}Value1}}}{6}
         {Value2}{{{\color{numb}Value2}}}{6}
         {:}{{{\color{punct}{:}}}}{1}
         {,}{{{\color{punct}{,}}}}{1}
         {\{}{{{\color{delim}{\{}}}}{1}
         {\}}{{{\color{delim}{\}}}}}{1}
         {[}{{{\color{delim}{[}}}}{1}
         {]}{{{\color{delim}{]}}}}{1},
}

\begin{document}

\title{The BIRAFFE2 Experiment.\\ %
  Study in \textbf{Bi}o-\textbf{R}eactions \textbf{a}nd \textbf{F}aces \textbf{f}or \textbf{E}motion-based Personalization %
  for AI Systems}

\author{\hspace{-1.3em} Krzysztof~Kutt
        \and Dominika~Dr\k{a}\.zyk\institute{Jagiellonian University, Poland, email: krzysztof.kutt@uj.edu.pl, dominika.a.drazyk@gmail.com}
        \and Maciej Szel\k{a}\.zek\institute{AGH University of Science and Technology, Poland, email: maciej.szelazek@agh.edu.pl}
        \and Szymon~Bobek
        \and Grzegorz~J.~Nalepa\institute{Jagiellonian University, Poland, email: szymon.bobek@uj.edu.pl, grzegorz.j.nalepa@uj.edu.pl}
}

\maketitle
\bibliographystyle{ecai}

\begin{abstract}
  The paper describes BIRAFFE2 data set, which is a result of an affective computing experiment conducted between 2019 and 2020, that aimed to develop computer models for classification and recognition of emotion.
  Such work is important to develop new methods of natural Human-AI interaction.
  As we believe that models of emotion should be personalized by design, we present an unified paradigm allowing to capture emotional responses of different persons, taking individual personality differences into account.
  We combine classical psychological paradigms of emotional response collection with the newer approach, based on the observation of the computer game player.
  By capturing ones psycho-physiological reactions (ECG, EDA signal recording), mimic expressions (facial emotion recognition), subjective valence-arousal balance ratings (widget ratings) and gameplay progression (accelerometer and screencast recording), we provide a framework that can be easily used and developed for the purpose of the machine learning methods. 
\end{abstract}

\section{INTRODUCTION}
\label{sec:intro}


Affective Computing (AfC)~\cite{picard1997affective}~-- an interdisciplinary field of study regarding human emotions~-- is to large extent built upon the assumption that we are able to precisely detect, label and manipulate emotional responses of agents. Therefore the proper understanding and modeling of this complex phenomena~\cite{handbookofemotions,calvo2015}, as well as maintaining ingenious experimental setup to do so, is a crucial determinant of success in this field. The setup should require conditions where human subjects are exposed to specific emotion evoking stimuli, and furthermore their reactions are somehow measured.

In our work we assume that the measurement of bodily reactions to stimuli can serve as proper foundation for emotion recognition (James-Lange approach~\cite{james1884what}). We also use a representation of affective data which is common in many experiments in psychology and human-computer interaction, i.e. the two dimensional Valence/Arousal space. We aim to continue a longer effort in building enhanced AfC models, previously presented on the AfCAI workshop, as well as later on in~\cite{gjn2019fgcs}. To make a measurement context as ecological as possible, we use wearable devices as well as other sensors which are possibly non-intrusive to the subjects. We also assume that the data processing should be possible offhand~-- with the use of devices that users have with them, e.g. mobile phones. Our recent results in this regards are summarized in~\cite{gjn2019sensors}.

The original contribution of the paper is the report on a an affective computing experiment conducted between 2019 and 2020, that aimed to develop computer models for classification and recognition of emotion.
The experiment resulted in the creation of publically developed dataset called BIRAFFE2.
We believe that this work might be an important step to develop new methods of natural Human-AI interaction.
The rest of the paper is organized as follows.
In Section~\ref{sec:motiv} we inroduce the motivation for our work and the context of the previous experiment in this area.
Then in Section~\ref{sec:bir2} we discuss our experimental methodology.
The structure of the resulting dataset is described in Section~\ref{sec:dataset}.
The paper ends with a summary and plans for future work in Section~\ref{sec:future}.

\section{MOTIVATION}
\label{sec:motiv}

In order to fully exploit advantages of user's emotion detection modules, whether in games or in apps, the recognition and classification of emotional states must be personalised.
By ``personalised'' we mean adjusted to ones personality traits and customs.
To implement that kind of tailoring, the appropriate variety of information about the user must be collected.
In order to verify these assumptions and to gather data for further development of the framework (for summary of the history of our approach see: \cite{gjn2019fgcs,gjn2019sensors}) we constructed a combined experimental paradigm.
First, in the ``classical approach'' part, we presented stimuli to the subjects and collected their answers: both with questionnaires and physiological signals measurement.
Secondly, within the ``ecological approach'', we embedded the reaction measurement in specific context of simple computer games.
When the player is loosing in the game, and the system detect the increased intensity of the her/his reaction, it is crucial for the model to interpret the context of such change~-- only by the contextual information coming from the game progression, we are able to tell whether the intensification was a collateral of anger or joy.
Knowing its contextual origin, we can also easily prevent these specific changes from happening again, or on the contrary~-- elicit detected and labeled emotions once again.
As such, games offer a perfect opportunity for putting the human in the feedback loop with the computer system (also called \emph{affective loop})~\cite{hook2008affective}.

The paradigm presented in this paper is the continuation of our previous work on the BIRAFFE1 experiment presented in~\cite{kkt2019afcai}.
In the current work, a number of improvements were introduced, drawing on the conclusions from the BIRAFFE1 study (details are presented in Section~\ref{sec:bir2}):
\begin{enumerate}
\item We improved the affect assessment widget.
\item We have enhanced the stimulus selection: it currently covers a wider area in the Valence-Arousal space, is more randomized and contains no erotic stimuli.
\item We have developed new custom computer games designed to arouse players' emotions -- unlike BIRAFFE1, here the games focus on a limited number of mechanics to evoke a limited set of emotions, which should make it easier to analyze game logs and draw conclusions.
\item We used the GEQ questionnaire to assess the involvement in the game, as well as asked about previous experiences with games to allow more accurate analysis of the emotions in games.
\item We extended the range of the measured responses by the introduction of accelerometer from game pad.
\item EDA and ECG electrodes placement was changed to overcome issues identified in previous experiment.
\item Several small improvements were also made, e.g., face photos are now taken with higher frequency, the screencast is recorded during game sessions.
\end{enumerate}



\section{METHODOLOGY}
\label{sec:bir2}

\subsection{Participants}

103 participants (33\% female) between 18 and 26 (\emph{M} = 21.63, \emph{SD} = 1.32) took part in the study. Information about recruitment was made available to students of the Artificial Intelligence Basics course at AGH UST, Krak\'{o}w, Poland. Although participation was not an obligatory part of the course, one could get bonus points for a personal participation or invitation of friends.


\subsection{Questionnaires}
\label{sec:questionnaires}

First, the paper-and-pen Polish adaptation~\cite{ptp1998neoffi} of the NEO Five Factor Inventory~\cite{costa1992neoffi} consisting of 60 self-descriptive statements evaluated on 1-5 scale (1 -- strongly disagree; 5 -- strongly agree) was used to measure the Big Five personality traits.

Second, our own paper-and-pen Polish translation of The Game Experience Questionnaire (GEQ) Core Module~\cite{ijsselsteijn2013geq} was used to measure players' feelings during the game session. The module consists of 33 items, ranked on 0-4 scale (0 -- not at all; 4 -- extremely). Items were arranged in seven components in original version: Competence, Sensory and Imaginative Immersion, Flow, Tension/Annoyance, Challenge, Negative affect and Positive affect.
The questionnaire has been used in many game studies~\cite{johnson2018pensgeq,law2018geq}. However, the 7-factor structure has not been confirmed by anyone.
In~\cite{johnson2018pensgeq} revised version (GEQ-R) was proposed.
Tension/Annoyance, Challenge and Negative affect were merged into Negativity, leading to 5-components solution.

Finally, our own simple questionnaire was used to measure gaming experience. It consists of two questions:
(1) ``Over the past year I have played computer / mobile / video games:''
(2) ``In the period of my most intense interest in computer / mobile / video games, I played:''.
Both were answered by selecting one of the five possible answers:
(a) daily or almost daily,
(b) several times a week,
(c) several times a month,
(d) several times a year,
(e) not at all.
There was also a space for leaving comments on the experiment. 


\subsection{Stimuli Selection}
\label{sec:stimuli}

\begin{figure}[htp]
  \centering
  \includegraphics[width=.4\textwidth]{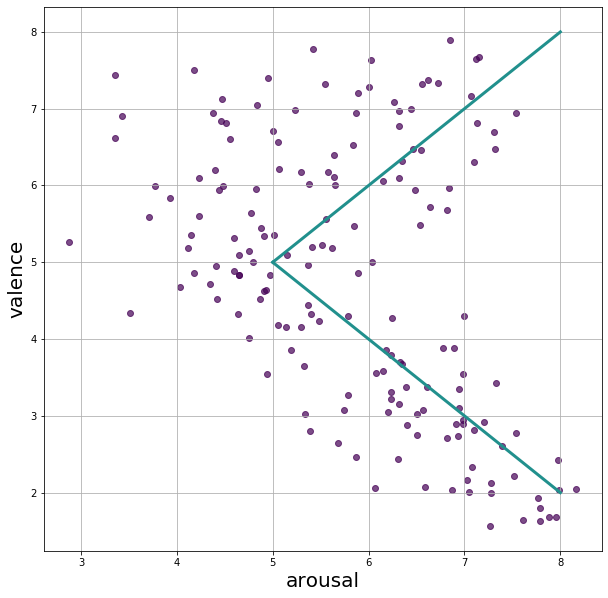}
  \caption{Trends in the IADS stimuli ratings.}
  \label{fig:iads2-trends}
\end{figure}

Standardized emotionally-evocative images and sounds from IAPS~\cite{lang2008iaps} and IADS~\cite{bradley2007iads} datasets were used as stimuli, each characterized by its coordinates in the Valence-Arousal space. The analysis of IADS and IAPS scores revealed the following trend: arousal score increases as the valence score strives for it's positive or negative extreme (Figure~\ref{fig:iads2-trends}). 

For the purpose of the experiment, we divided the stimuli into three groups according to their arousal and valence index: + (positive valence and high arousal), 0 (neutral valence and medium arousal), -- (negative valence and high arousal). Erotic stimuli were excluded from the blind selection, due to the risk of creating weird or disgusting combinations (e.g. picture of kid or snake is paired with the erotic sound), not intended by the aim of this study. 

The stimuli set for each participant was generated by random sampling without replacement and formed nine conditions, each consisting of 13 stimuli:
\begin{itemize}
\item three consisting only of pictures: \emph{p+}, \emph{p0}, \emph{p--},
\item three consisting only of sounds: \emph{s+}, \emph{s0}, \emph{s--},
\item three, where pictures were paired with sounds: 
+ picture with + sound (\emph{p+s+}),
0 picture with 0 sound (\emph{p0s0}),
-- picture with -- sound (\emph{p--s--}).
\end{itemize}
Conditions were mixed during the presentation, which was divided into two sessions (17.5 min each) and separated by the game session.


\subsection{Emotion Evaluation Widget}
\label{sec:widget}

\begin{figure*}[htp]
  \begin{center}
        \includegraphics[width=.3\textwidth]{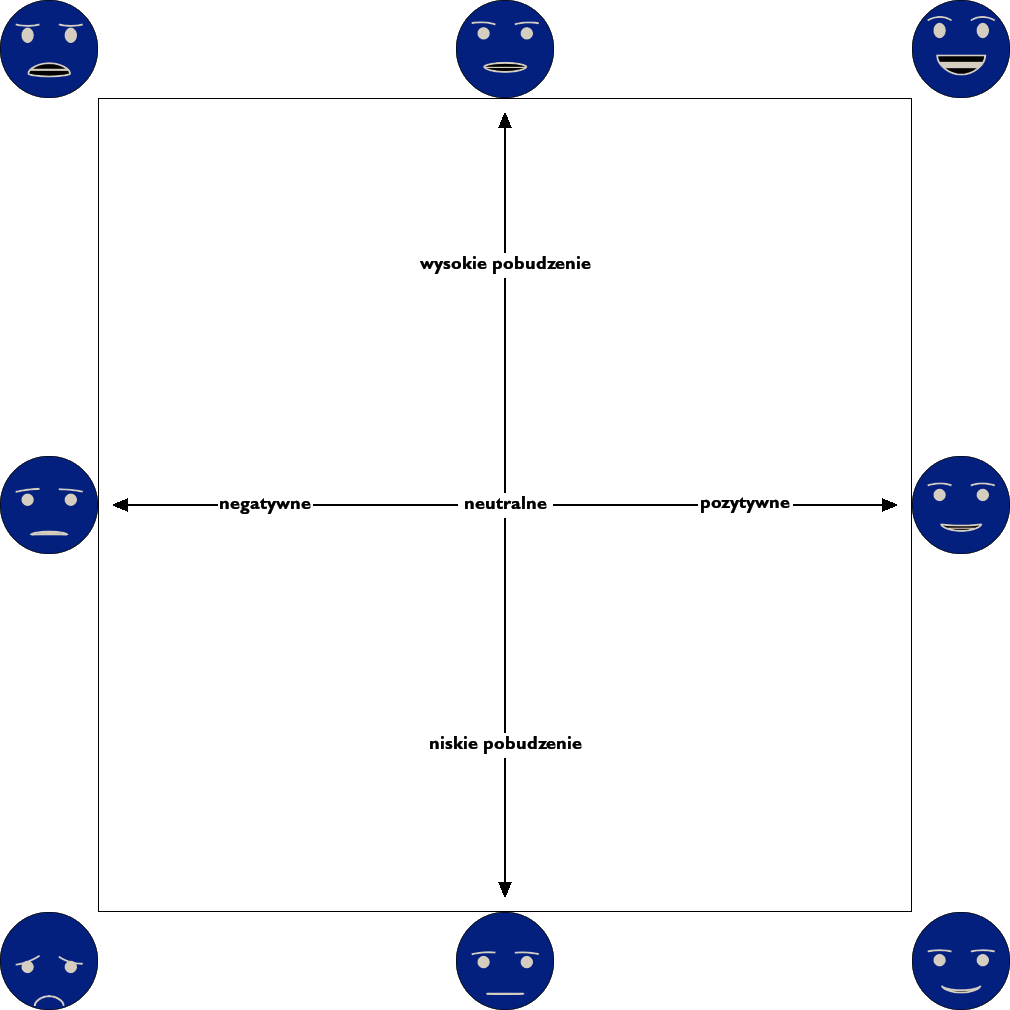}
    ~\hspace{9em} 
        \includegraphics[width=.3\textwidth]{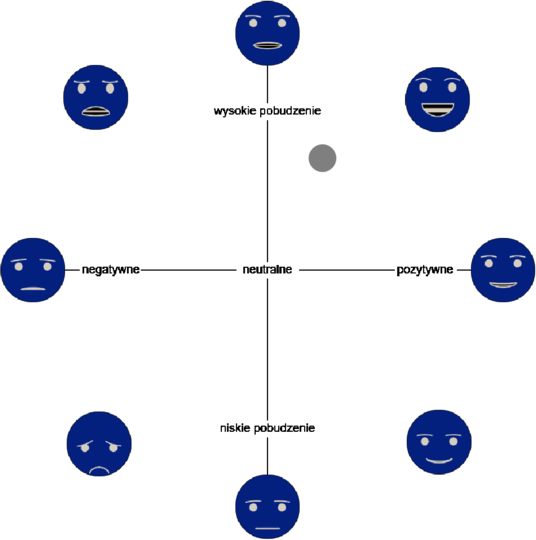}
        \caption{Current and previous~\cite{kkt2019afcai} versions of the \emph{Valence-arousal faces} widget (pictures presented with a negative filter).}
    \label{fig:widget}
  \end{center}    
\end{figure*}

Emotional assessment was carried out using the \emph{Valence-arousal faces} widget controlled by a left joystick on a gamepad. Widget was adapted from our previous experiment~\cite{kkt2019afcai}, with the following improvements:
\begin{itemize}
\item Emoticons placed as hints were moved outside the selection area. Also, the border of the selection area was introduced. In the previous version the subjects often chose the location of the smiley as their answer. Now, there is no possibility to put selection marker on them.
\item The selection marker changes color to indicate that there is only half a second left for the rating. In the previous version there was no information about the remaining time.
\item The returned valence and arousal scores are now within a range of $[1, 9]$, so they are within the same range as the assessments in IAPS and IADS. In the previous version, they were in the $[-1, 1]$ range, which required conversion of values before the analysis started.
\end{itemize}
To compare current and previous version of \emph{Valence-arousal faces} widget see Figure~\ref{fig:widget}.
Both are presented in Polish, as in studies.
X~axis has labels ``negative'', ``neutral'', ``positive'', while Y axis has labels: ``high arousal'' and ``low arousal''.


\subsection{Games}

\begin{figure}[htp]
  \centering
  \includegraphics[width=.35\textwidth]{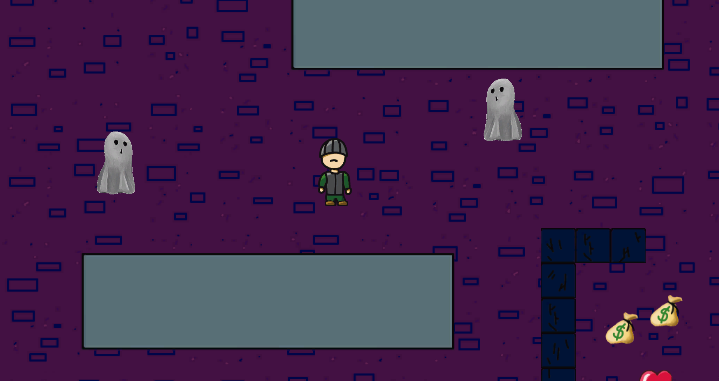}
  \caption{\emph{Room of the Ghosts} gameplay.}
  \label{fig:game1}
\end{figure}

\begin{figure}[htp]
  \centering
  \includegraphics[width=.35\textwidth]{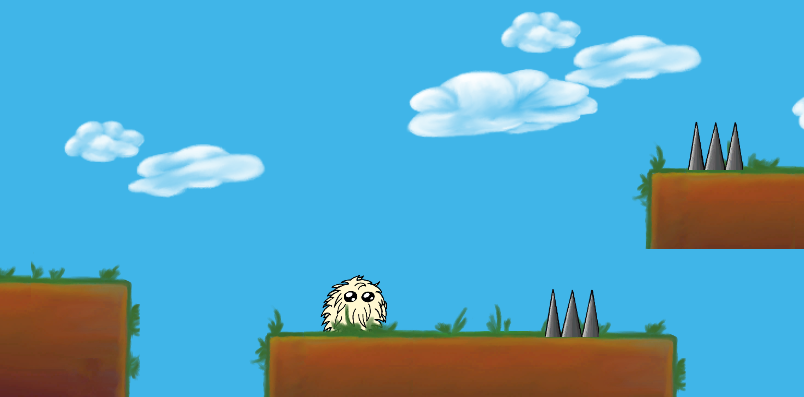}
  \caption{\emph{Jump!} gameplay.}
  \label{fig:game2}
\end{figure}

\begin{figure}[htp]
  \centering
  \includegraphics[width=.35\textwidth]{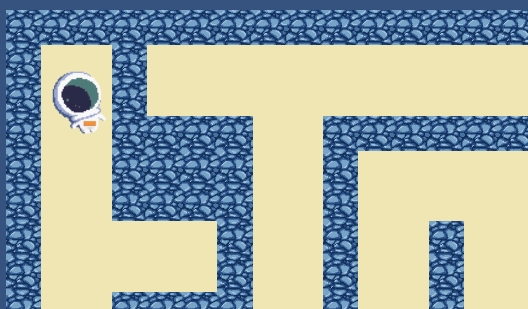}
  \caption{\emph{Labyrinth} gameplay.}
  \label{fig:game3}
\end{figure}

Three affective games developed by our team~\cite{zuchowska2020eng} were used during the study. All of them were controlled by a gamepad and produced game log CSV files. They have been designed with the emphasis on differentiating the levels of difficulty:

\begin{itemize}
\item \emph{Room of the Ghosts}:
    The goal: pass through a series of rooms and defeat the arriving ghosts (see Figure~\ref{fig:game1}).
    Difficulty: very easy, achieved by the following implementations: collider for protagonist is smaller than his real model -- it removes the feeling of being hit before the projectile hits the player; the protagonist's weapon that can shoot more often and faster than the opponents' weapons.

\item \emph{Jump!}:
    The goal: reach the end of the path by jumping on the platforms and avoiding obstacles (see Figure~\ref{fig:game2}).
    Difficulty: hard and frustrating, achieved by the following implementations: colliders are too big -- player can get hit by trap before s/he touches it with the model; movement is clunky, and there are several traps, i.e. invisible blocks, which increases the confusion and irritation in player; each time the protagonist dies, the background music is getting less pleasant (the pitch and distort levels of music playing in background increases by 0.07).

\item \emph{Labyrinth}:
    The goal: walk the protagonist through the labyrinth (see Figure~\ref{fig:game3}).
    Difficulty: optimal, achieved by the following implementations: the colliders have been adjusted to not hit the walls too often and to make the movement smooth; the protagonist control is natural and predictable.
\end{itemize}

\subsection{Hardware}

\begin{figure}[htp]
  \centering
  \includegraphics[width=.4\textwidth]{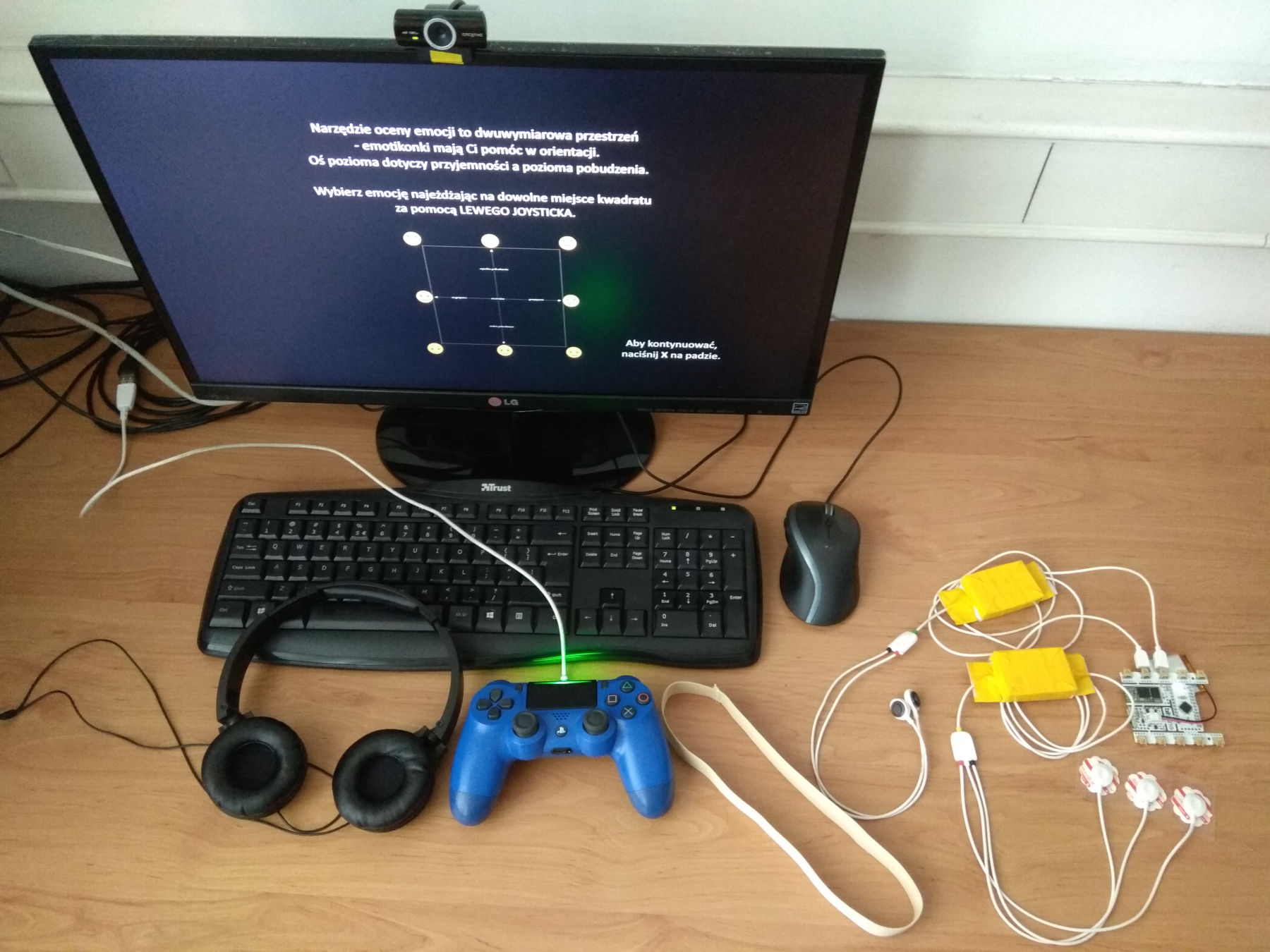}
  \caption{Research setup: 23'' LCD screen, headphones, external web camera, gamepad, band for EDA electrodes and Bitalino with 3-leads ECG and 2-leads EDA. Keyboard and mice were used only by the researcher to start the protocol.}
  \label{fig:setup}
\end{figure}

Experimental setup consists of (see Figure~\ref{fig:setup}):
\begin{itemize}
\item Full HD 23'' LCD screen,
\item PC (processor: Intel Core i5-8600K, graphic card: MSI GeForce GTX 1070, 16 GB RAM) running under the 64-bit Windows 10 1909 Education,
\item External web camera Creative Live! Cam Sync HD 720p, 
\item Gamepad Sony PlayStation DualShock 4,
\item Pioneer SE-MJ503 headphones,
\item BITalino (r)evolution kit\footnote{See: \url{https://bitalino.com/en/}.} with 3-leads ECG and 2-leads EDA sensors,
\item Body-coloured band to hold the EDA electrodes. 
\end{itemize}

\subsection{Software}

The procedure was running under the Python 3.8, written with PsychoPy 3.2.4 library~\cite{peirce2019psychopy}.
Python code controlled the execution of the whole protocol, i.e. stimuli presentation, screencast recording (using OBS Studio\footnote{See: \url{https://obsproject.com/}.} software), photos taking, games' start and end management.

Physiological signals were gathered using BITalino (r)evolution kit, as it is the most promising of cheap mobile hardware platforms
(for comparison see~\cite{afcsensors-icaisc2018}). Electrocardiogram recording was implemented using the classical 3 leads montage with electrodes placed below the collarbones (V-- and reference) and below the last rib on the left side of the body (V+).
EDA signal was gathered by 2 leads placed on the forehead (placement as good as classical palmar location~\cite{dooren2012gsr16locations}, with no side effects related to gamepad held by the subjects). EDA electrodes were hidden under the body-coloured band, to not interfere with the facial emotion recognition process conducted by the API in the later experimental stages, and to provide tight and stable contact of the EDA electrode and skin.
Both signals were probed with 1000 Hz sampling rate.


\subsection{Procedure}

Two research stands were prepared in the same room, located next to the opposite walls, so that the subjects sat back to back. Each participant was seated in front of a monitor and provided with the consent and short information about the whole experiment. Throughout the whole procedure, the investigator remained at his desk, rear-facing to the subjects, to reduce the Hawthrone effect.

Subjects filled out the NEO-FFI inventory and were connected to measuring devices, with the headphones set up, and a gamepad given to hand. Computer protocol consisted of five phases:
\begin{enumerate}
\item Baseline signals recording (1 minute),
\item Instructions and training (approx. 5 minutes),
\item First part of stimuli presentation and rating (17.5 minutes): each presentation lasted 6 seconds\footnote{Each sound in IADS set lasts 6 seconds.}, followed by 6 seconds for affective rating and 6 second ISI\footnote{It results in 18 seconds between stimuli onset, which is enough for observing reactions in the ECG and GSR signals.}. Participants were instructed to navigate the procedure via gamepad.
\item Games session (up to 15 minutes in total): each game had a 5 min time limit, after which it turned itself off. After the completion of one game, another automatically switched on.
\item Second part of stimuli presentation and rating (17.5 minutes).
\end{enumerate}
After the computer protocol, subjects filled out three GEQ questionnaires (one for each game) and gaming experience questionnaire.

ECG and GSR signal, as well as gamepad accelerometer and gyroscope readings, were collected continuously during the whole experiment.
Facial photos were taken every 250 milliseconds. A screencast was recorded during the game session, in case for the need to fill in the missing information in game logs after the experiment. The whole protocol lasted up to 75 minutes.


\section{THE BIRAFFE2 DATASET}
\label{sec:dataset}

BIRAFFE2 dataset is available to download at Zenodo\footnote{See: \url{https://dx.doi.org/10.5281/zenodo.3865859}.}.  
\\*
It consists of a metadata file and 7 archives of the following data collections:
\begin{description}
\item[\texttt{BIRAFFE2-metadata.csv}] contains a summary of each participant:
    age, sex, personality profile, GEQ results and information about subsets available for given person (whether there is a \texttt{BioSigs}, \texttt{Screencast}, \ldots file available for the person),
\item[\texttt{BIRAFFE2-biosigs.zip}] contains biosignals (ECG and GSR),
\item[\texttt{BIRAFFE2-gamepad.zip}] contains accelerometer and gyroscope recordings,
\item[\texttt{BIRAFFE2-games.zip}] contains logs from the games,
\item[\texttt{BIRAFFE2-photo.zip}] contains face emotions description calculated by MS Face API,
\item[\texttt{BIRAFFE2-photo-full.zip}] contains all information available in \texttt{BIRAFFE2-photo.zip} but also other face-related values recognized by MS Face API, e.g. recognized age, whether the person wears glasses, what is the color of the hair, 
\item[\texttt{BIRAFFE2-procedure.zip}] contains a log of all the stimuli presented to a given user (timestamp of the stimuli presentation, condition, widget, stimuli ID...),
\item[\texttt{BIRAFFE2-screencast.zip}] contains screencasts from games.
\end{description}

All files have Unix timestamps which can be used for synchronization between different subsets.
Detailed low-level specification of all values is provided in Sect.~\ref{sec:metadata}-\ref{sec:screencast}.

BIRAFFE2 dataset consists of data gathered from 102 out of 103 participants.
Unfortunately, during the protocol for subject 723 there were problems with the hard drive and all data was lost, except the paper-and-pen NEO-FFI and GEQ questionnaires.
Also, some smaller issues occurred for a subset of participants,
e.g. game crashed, Bluetooth signal was lost, electrode contact was poor.
We have published also the incomplete records (see Tab.~\ref{tab:stats}), as in many analysis only selected of the subsets will be used and it will not be the problem. Missing values in all files are represented by \texttt{NaN}.

\begin{table}
  \centering
  \caption{Size of subsets collected (out of 103 participants)}
  \label{tab:stats}
  \begin{tabular}{|r|l|}
  \hline
  GEQ \& NEO-FFI & 103 \\
  BioSigs & 102 \\
  Gamepad & 102 \\
  Game 1 logs & 102 \\
  Game 2 logs & 101 \\
  Game 3 logs & 87 \\
  Photos & 102 \\
  Procedure & 102 \\
  Screencast & 92 \\ \hline
  \end{tabular} 
\end{table}


\subsection{\texttt{BIRAFFE2-metadata.csv}}
\label{sec:metadata}

Each line of this file represents one subject and consists of the following values:
\begin{lstlisting}[numbers=none]
ID;AGE;SEX;NEO-FFI;GEQ;BIOSIGS;GAMEPAD;GAME-1;GAME-2;GAME-3;PHOTOS;PROCEDURE;SCREENCAST;OPENNESS;CONSCIENTIOUSNESS;NEUROTICISM;AGREEABLENESS;EXTRAVERSION;GAME-EXO-PAST-YEAR;GAME-EXP-MOST-INTENSE;GEQ-1-COMPETENCE-2013;GEQ-1-IMMERSION-2013;GEQ-1-FLOW-2013;GEQ-1-TENSION-2013;GEQ-1-CHALLENGE-2013;GEQ-1-NEGATIVE-AFFECT-2013;GEQ-1-POSITIVE-AFFECT-2013;GEQ-1-POSITIVE-AFFECT-2018;GEQ-1-NEGATIVITY-2018;GEQ-1-COMPETENCE-2018;GEQ-1-FLOW-2018;GEQ-1-IMMERSION-2018;GEQ-2-COMPETENCE-2013;GEQ-2-IMMERSION-2013;GEQ-2-FLOW-2013;GEQ-2-TENSION-2013;GEQ-2-CHALLENGE-2013;GEQ-2-NEGATIVE-AFFECT-2013;GEQ-2-POSITIVE-AFFECT-2013;GEQ-2-POSITIVE-AFFECT-2018;GEQ-2-NEGATIVITY-2018;GEQ-2-COMPETENCE-2018;GEQ-2-FLOW-2018;GEQ-2-IMMERSION-2018;GEQ-3-COMPETENCE-2013;GEQ-3-IMMERSION-2013;GEQ-3-FLOW-2013;GEQ-3-TENSION-2013;GEQ-3-CHALLENGE-2013;GEQ-3-NEGATIVE-AFFECT-2013;GEQ-3-POSITIVE-AFFECT-2013;GEQ-3-POSITIVE-AFFECT-2018;GEQ-3-NEGATIVITY-2018;GEQ-3-COMPETENCE-2018;GEQ-3-FLOW-2018;GEQ-3-IMMERSION-2018
\end{lstlisting}
where\footnote{Note that not all values are described in details, as some of them are obvious.}:
\begin{itemize}
\item \textbf{\texttt{ID}} -- a randomly assigned subject ID from range \{100,999\}. It is used to identify all subject-related files as filenames. Filenames are created according to the format \texttt{SUBxxx-yyyy}, where \texttt{xxx} is the ID, and \texttt{yyyy} is the data type identifier (e.g. \texttt{BioSigs}, \texttt{Gamepad}),
\item \textbf{\texttt{NEO-FFI;GEQ;BIOSIGS;GAMEPAD;GAME-1;GAME-2; GAME-3;PHOTOS;PROCEDURE;SCREENCAST}} -- information about subsets available for given person, i.e. whether there is a \texttt{BioSigs} file, \texttt{Gamepad} file, etc. available for the person (\texttt{Y} or \texttt{NaN}). \texttt{GAME-X} columns inform about availability of game logs for level \texttt{X}. The NEO-FFI and GEQ columns only indicate whether there are questionnaire results in the following columns,
\item \textbf{\texttt{OPENNESS;CONSCIENTIOUSNESS;EXTRAVERSION; AGREEABLENESS;NEUROTICISM}} -- five personality traits calculated from raw NEO-FFI results; values represent ten scores, i.e. the possible values are in \{1,2,3,\ldots,10\} set and represent standard normal distribution with $M=5.5$ and $SD=2$. For further analyses they can be transformed to low (1-3), medium (4-6) and high (7-10) trait levels,
\item \textbf{\texttt{GAME-EXO-PAST-YEAR;GAME-EXP-MOST-INTENSE}} -- answers from our own simple questionnaire for gaming experience measurement described in Sect.~\ref{sec:questionnaires}. The possible values are \texttt{\{A,B,C,D,E\}},
\item \textbf{\texttt{GEQ-X-Y-Z}} -- are calculated components values from GEQ questionnaires. X represents the game level (\texttt{\{1,2,3\}}),
Y -- the component name,
Z -- version of the GEQ (\texttt{2013} is the original work~\cite{ijsselsteijn2013geq}, while \texttt{2018} is the revised version~\cite{johnson2018pensgeq}). The values are ranging from 0 (not at all) to 4 (extremely) for each factor.
\end{itemize}


\subsection{\texttt{BIRAFFE2-biosigs.zip}}
\label{sec:biosigs}

Each \texttt{SUBxxx-BioSigs.csv} file represents one subject and consists of one line per each \emph{sensor recording}. Values were recorded with 1~\si{\kilo\hertz} frequency. The fields contained in each line are:
\begin{lstlisting}[numbers=none]
TIMESTAMP;ECG;GSR
\end{lstlisting}
where:
\begin{itemize}
\item \textbf{\texttt{ECG}} -- signal (units: \si{\milli\volt}) gathered by BITalino, after units transformation recommended by BITalino Sensor Datasheet\footnote{See: \url{https://bitalino.com/datasheets/ECG_Sensor_Datasheet.pdf}.}, low-pass filtering in 35 \si{\hertz} and baseline removal performed using Python library for biosignal processing~\cite{heartpy}.
\item \textbf{\texttt{GSR}} -- signal (units: \si{\micro\siemens}) gathered by BITalino, after units transformation recommended by BITalino Sensor Datasheet\footnote{See: \url{https://bitalino.com/datasheets/EDA_Sensor_Datasheet.pdf}.} and low-pass filtering (in range between 0.5 to 50 \si{\hertz} depending on the noise level in the individual file) performed using Python library for biosignal processing~\cite{heartpy}.
\end{itemize}
The signals were recorded using the Python library for BITalino\footnote{See: \url{http://bitalino.com/pyAPI/}.}.
Due to the instability of the bluetooth connection,
and in the absence of handling such a situation in the library,
the timestamps after series\footnote{Errors are considered to occur in the series if the interval between consecutive error entries in the \textbf{Procedure} log (see Sect.~\ref{sec:procedure}) was less than 0.05~\si{\second}.} of several connection errors
cannot be considered as fully reliable.
This will be further investigated by our team.


\subsection{\texttt{BIRAFFE2-gamepad.zip}}

Each \texttt{SUBxxx-gamepad.csv} file represents one subject and consists of one line per each \emph{gamepad recording}. The values were recorded as quickly as they were transmitted through the USB interface, with an average frequency of 250~\si{\hertz}. The fields contained in each line are: 
\begin{lstlisting}[numbers=none]
TIMESTAMP;GYR-X;GYR-Y;GYR-Z;ACC-X;ACC-Y;ACC-Z
\end{lstlisting}
where \texttt{GYR-} are gyroscope readings,
while \texttt{ACC-} are accelerometer values.
Note that the gyroscope values represent the \emph{position} of the gamepad,
not the angular rate\footnote{As in the DS4Windows library, which code was used as the base for our Python data acquisition code. See: \url{https://github.com/Jays2Kings/DS4Windows}.}:
\begin{itemize}
\item \textbf{\texttt{GYR-X}} -- right side of the gamepad upward,
\item \textbf{\texttt{GYR-Y}} -- buttons and joysticks panel upward,
\item \textbf{\texttt{GYR-Z}} -- audio port upward / light bar downward,
\item \textbf{\texttt{ACC-X}} -- yaw counter-clockwise,
\item \textbf{\texttt{ACC-Y}} -- pitch upward,
\item \textbf{\texttt{ACC-Z}} -- roll left side of gamepad down.
\end{itemize}



\subsection{\texttt{BIRAFFE2-games.zip}}

Five JSON log files are created for each subject.
Note that the file was created when the level has started.
If the game has crashed in a given level (which sometimes happened),
the subsequent levels were not started and the subject was returned to the stimulus session.
This means that in some cases files for may be missing,
e.g. files for level 3 if the game crashed at the second level.

\paragraph{\texttt{SUBxxx-Level01\_Log.json}} contains a log from the \emph{Room of the Ghosts} composed of information collected at multiple points in time about the current position and status of the user.
It is a repeated pattern of the following structure:
\begin{lstlisting}[language=json]
[
  {
    "x": Float,  # current position on the map
    "y": Float,
    "deathCount": Int,
    "shootsCounter": Int,  # both may be used to
    "hitCounter": Int,     # calculate the accuracy
                           # of the player
    "money": Int,  # current score (death: -5)
    "collectedMoney": Int,   # how many pickups
    "collectedHealth": Int,  # were collected
    "health": Int,
    "timestamp": Int,  # current Unix timestamp
    "idOfSound": String,  # ID of IADS sound played
                          # in the background
    "timestampOfSound": String,  # time when sound
                                 # was started
    "xMin": Float,  # visible area of the map
    "yMin": Float,
    "xMax": Float,
    "yMax": Float
  },
  ...
]
\end{lstlisting}

\paragraph{\texttt{SUBxxx-Level01\_MapLog.json}} contains information about the dynamic environment in the \emph{Room of the Ghosts}. It is composed of series of three subsequent lists -- each describing current position of existing ghosts, money bags and health pickups:
\begin{lstlisting}[language=json]
[
  {
    "Ghosts": [
      {
        "x": Float,  # current position
        "y": Float,
        "timestamp": Int  # Unix timestamp when the
                          # position was recorded
      },
      ...  # each ghost has a separate entry
  },
  {
    "Money bags": [
      ...  # the list is the same as for ghosts
    ]
  },
  {
    "Health pickups": [
      ...  # the list is the same as for ghosts
    ]
  },
  ...
]
\end{lstlisting}

\paragraph{\texttt{SUBxxx-Level02\_Log.json}} is a log from the \emph{Jump!}, as the log from the first level, it is composed of a repeated pattern collected at multiple points in time:
\begin{lstlisting}[language=json]
[
  {
    "x": Float,  # current position
    "y": Float,
    "deathCount": Int,
    "distortionLevel": Float,  # current parameters
    "pitchLevel": Float,       # of the sound 
    "timestamp": Int,
    "idOfSound": String,  # as in the first level
    "timestampOfSound": String,
    "xMin": Float,  # visible area of the map
    "yMin": Float,
    "xMax": Float,
    "yMax": Float
  },
  ...
]
\end{lstlisting}

\paragraph{\texttt{SUBxxx-Level02\_BlockEvents.json}} contains the information about the dynamic blocks of the \emph{Jump!} game world. The design is analogous to \texttt{SUBxxx-Level01\_MapLog.json}:
\begin{lstlisting}[language=json]
[
  {
    "Invisible blocks": [
      {
        "x": Float,
        "y": Float,
        "timestamp": Int
      },
      ...  # each block has a separate entry
  },
  {
    "Falling blocks": [
      ...  # the list is the same as above
    ]
  },
  ...
]
\end{lstlisting}

\paragraph{\texttt{SUBxxx-Level03\_Log.json}} is a log from the \emph{Labyrinth}, as the previous logs, it consists of often sampled structure:
\begin{lstlisting}[language=json]
[
  {
    "x": Float,  # current position
    "y": Float,
    "deadEnds": Int,  # counts how many times
                      # the player was off the
                      # correct path
    "wasOnCorrectPath": Bool,
    "timestamp": Int,
    "idOfSound": String,
    "timestampOfSound": String,
    "xMin": Float,
    "yMin": Float,
    "xMax": Float,
    "yMax": Float
  },
  ...
]
\end{lstlisting}

There are also three files containing static map of each level (same for each subject): \textbf{\texttt{Level01\_StaticMap.json}}, \textbf{\texttt{Level02\_StaticMap.json}}, \textbf{\texttt{Level03\_StaticMap.json}}.
They are located in the root directory of the subset.
Each of them consists of a list of position of all (squared) blocks building the maps:
\begin{lstlisting}[language=json]
[
  { "x": Float, "y": Float, "z": Float },  # 1st
  { "x": Float, "y": Float, "z": Float },  # 2nd
  ...
]
\end{lstlisting}


\subsection{\texttt{BIRAFFE2-photo.zip}}
\label{sec:photo}

Each \texttt{SUBxxx-Face.csv} file represents one subject and consists of one line per each \emph{photo taken}.
Raw photos are not available due to privacy reasons.
File consists of values calculated by MS Face API with \emph{recognition\_02} model. Photos were taken with 4 \si{\hertz} frequency during games and during stimuli presentation (every 15 frames at 60 \si{fps}). Photos were not taken while the subject was responding on the widget. When no face was recognized or two faces were found (the second was the experimenter face) \texttt{NaN} value was used.
\begin{lstlisting}[numbers=none]
GAME-TIMESTAMP;FRAME-NUMBER;IADS-ID;IAPS-ID;ANGER;CONTEMPT;DISGUST;FEAR;HAPPINESS;NEUTRAL;SADNESS;SURPRISE
\end{lstlisting}
where:
\begin{itemize}
\item \textbf{\texttt{GAME-TIMESTAMP}} -- Unix timestamp available only during the game (\texttt{NaN} value during the stimuli presentation),
\item \textbf{\texttt{FRAME-NUMBER}} -- Index of the photo within the context of the stimuli presentation (\texttt{NaN} value during the games), measured in frames since the beginning  of the stimuli presentation: $-1$ for pre-stimulation photo, $0$ for photo in the moment when stimuli appears, $15$ for the next photo ($1/4$ \si{\second} later), up to $345$ (frame $360$ = $6$ \si{\second} = time when stimuli disappears),
\item \textbf{\texttt{IADS-ID;IAPS-ID}} -- IADS/IAPS IDs of stimuli (see Sect.~\ref{sec:procedure}),
\item \textbf{\texttt{ANGER;CONTEMPT;DISGUST;FEAR;HAPPINESS; NEUTRAL;SADNESS;SURPRISE}} -- probability distribution of eight emotions calculated by MS Face API (all values sum up $1$). It is important to note that this distribution is highly skewed to the \texttt{NEUTRAL} emotion, having values close to 1 in that emotion and values close to zero in the rest of them.
\end{itemize}


\subsection{\texttt{BIRAFFE2-photo-full.zip}}

\texttt{SUBxxx-Face.csv} files are analogous to the \texttt{BIRAFFE-photo.zip}, but with the full output from MS Face API. Besides the values described in Sect.~\ref{sec:photo},
they also have the following face-related values\footnote{See also MS Face API documentation at \url{https://westus.dev.cognitive.microsoft.com/docs/services/563879b61984550e40cbbe8d/operations/563879b61984550f30395236}.}:
\begin{lstlisting}[numbers=none]
FACEATTRIBUTES-ACCESSORIES;FACEATTRIBUTES-AGE;FACEATTRIBUTES-BLUR-BLURLEVEL;FACEATTRIBUTES-BLUR-VALUE;FACEATTRIBUTES-EXPOSURE-EXPOSURELEVEL;FACEATTRIBUTES-EXPOSURE-VALUE;FACEATTRIBUTES-FACIALHAIR-BEARD;FACEATTRIBUTES-FACIALHAIR-MOUSTACHE;FACEATTRIBUTES-FACIALHAIR-SIDEBURNS;FACEATTRIBUTES-GENDER;FACEATTRIBUTES-GLASSES;FACEATTRIBUTES-HAIR-BALD;FACEATTRIBUTES-HAIR-HAIRCOLOR-BLACK;FACEATTRIBUTES-HAIR-HAIRCOLOR-BLOND;FACEATTRIBUTES-HAIR-HAIRCOLOR-BROWN;FACEATTRIBUTES-HAIR-HAIRCOLOR-GRAY;FACEATTRIBUTES-HAIR-HAIRCOLOR-OTHER;FACEATTRIBUTES-HAIR-HAIRCOLOR-RED;FACEATTRIBUTES-HAIR-INVISIBLE;FACEATTRIBUTES-HEADPOSE-PITCH;FACEATTRIBUTES-HEADPOSE-ROLL;FACEATTRIBUTES-HEADPOSE-YAW;FACEATTRIBUTES-MAKEUP-EYEMAKEUP;FACEATTRIBUTES-MAKEUP-LIPMAKEUP;FACEATTRIBUTES-NOISE-NOISELEVEL;FACEATTRIBUTES-NOISE-VALUE;FACEATTRIBUTES-SMILE;FACEID;FACELANDMARKS-EYEBROWLEFTINNER-X;FACELANDMARKS-EYEBROWLEFTINNER-Y;FACELANDMARKS-EYEBROWLEFTOUTER-X;FACELANDMARKS-EYEBROWLEFTOUTER-Y;FACELANDMARKS-EYEBROWRIGHTINNER-X;FACELANDMARKS-EYEBROWRIGHTINNER-Y;FACELANDMARKS-EYEBROWRIGHTOUTER-X;FACELANDMARKS-EYEBROWRIGHTOUTER-Y;FACELANDMARKS-EYELEFTBOTTOM-X;FACELANDMARKS-EYELEFTBOTTOM-Y;FACELANDMARKS-EYELEFTINNER-X;FACELANDMARKS-EYELEFTINNER-Y;FACELANDMARKS-EYELEFTOUTER-X;FACELANDMARKS-EYELEFTOUTER-Y;FACELANDMARKS-EYELEFTTOP-X;FACELANDMARKS-EYELEFTTOP-Y;FACELANDMARKS-EYERIGHTBOTTOM-X;FACELANDMARKS-EYERIGHTBOTTOM-Y;FACELANDMARKS-EYERIGHTINNER-X;FACELANDMARKS-EYERIGHTINNER-Y;FACELANDMARKS-EYERIGHTOUTER-X;FACELANDMARKS-EYERIGHTOUTER-Y;FACELANDMARKS-EYERIGHTTOP-X;FACELANDMARKS-EYERIGHTTOP-Y;FACELANDMARKS-MOUTHLEFT-X;FACELANDMARKS-MOUTHLEFT-Y;FACELANDMARKS-MOUTHRIGHT-X;FACELANDMARKS-MOUTHRIGHT-Y;FACELANDMARKS-NOSELEFTALAROUTTIP-X;FACELANDMARKS-NOSELEFTALAROUTTIP-Y;FACELANDMARKS-NOSELEFTALARTOP-X;FACELANDMARKS-NOSELEFTALARTOP-Y;FACELANDMARKS-NOSERIGHTALAROUTTIP-X;FACELANDMARKS-NOSERIGHTALAROUTTIP-Y;FACELANDMARKS-NOSERIGHTALARTOP-X;FACELANDMARKS-NOSERIGHTALARTOP-Y;FACELANDMARKS-NOSEROOTLEFT-X;FACELANDMARKS-NOSEROOTLEFT-Y;FACELANDMARKS-NOSEROOTRIGHT-X;FACELANDMARKS-NOSEROOTRIGHT-Y;FACELANDMARKS-NOSETIP-X;FACELANDMARKS-NOSETIP-Y;FACELANDMARKS-PUPILLEFT-X;FACELANDMARKS-PUPILLEFT-Y;FACELANDMARKS-PUPILRIGHT-X;FACELANDMARKS-PUPILRIGHT-Y;FACELANDMARKS-UNDERLIPBOTTOM-X;FACELANDMARKS-UNDERLIPBOTTOM-Y;FACELANDMARKS-UNDERLIPTOP-X;FACELANDMARKS-UNDERLIPTOP-Y;FACELANDMARKS-UPPERLIPBOTTOM-X;FACELANDMARKS-UPPERLIPBOTTOM-Y;FACELANDMARKS-UPPERLIPTOP-X;FACELANDMARKS-UPPERLIPTOP-Y;FACERECTANGLE-HEIGHT;FACERECTANGLE-LEFT;FACERECTANGLE-TOP;FACERECTANGLE-WIDTH
\end{lstlisting}


\subsection{\texttt{BIRAFFE2-procedure.zip}}
\label{sec:procedure}

Each \texttt{SUBxxx-Procedure.csv} file represents one subject and consists of one line per each \emph{stimuli presentation}. The fields contained in each line are:
\begin{lstlisting}[numbers=none]
TIMESTAMP;ID;COND;IADS-ID;IAPS-ID;ANS-VALENCE;ANS-AROUSAL;ANS-TIME;EVENT
\end{lstlisting}
where:
\begin{itemize}
\item \textbf{\texttt{TIMESTAMP}} -- Unix timestamp when the stimuli appeared on the screen,
\item \textbf{\texttt{ID}} -- subject ID,
\item \textbf{\texttt{COND}} -- one of nine conditions as specified in Sect.~\ref{sec:stimuli} (\texttt{P+, P0, P--, S+, S0, S--, P+S+, P0S0, P--S--}), 
\item \textbf{\texttt{IADS-ID;IAPS-ID}} -- IADS/IAPS IDs of stimuli. Both IADS and IAPS datasets provide Valence/Arousal scores for each stimuli that can be used for further analyses (these values describe emotions that were evoked by the stimuli). Contact with the CSEA at University of Florida to obtain your own copy of the datasets for research\footnote{See: \url{https://csea.phhp.ufl.edu/media.html}.},
\item \textbf{\texttt{ANS-VALENCE;ANS-AROUSAL}} -- values in $[1;9]$ ranges indicating the point selected by the subject in the \emph{Valence-arousal faces} widget (see Sect.~\ref{sec:widget}),
\item \textbf{\texttt{ANS-TIME}} -- response time (0 is a moment when widget appeared on the screen); \texttt{NaN} indicates that the subject has not made any choice but left the default option,
\item \textbf{\texttt{EVENT}} -- information about going through the next procedure checkpoint (e.g. tutorial start, game session end) and about BITalino errors (see Sect.~\ref{sec:biosigs}).
\end{itemize}


\subsection{\texttt{BIRAFFE2-screencast.zip}}
\label{sec:screencast}

Each \texttt{SUBxxx-Screencast.mkv} file contains a \emph{screen recording} (1920x1080 resolution, 60 fps, h.264 codec) of game session for one subject.
The entire subset weights 12 GB and, due to its large size, it is available on request.


\section{SUMMARY AND FUTURE WORK}
\label{sec:future}

This paper describes the BIRAFFE2 data set collected in 2020 during the affective computing experiment, aiming to provide a framework for the enhanced models of emotion classification an recognition. The enhancement, founded by the variety of collected data types and behavioural features, is grounded in the belief, that the up-to-date emotion recognition models should be personalized by design. The construction of such models, as well as the experimental design used for data collection, should take the individual personality differences into account and provide a solution for its operationalization. We believe, that the data set described in this paper presents an important contribution that supports the development and replication of experiments in affective computing and Human-AI interaction.

Based on the data acquired in the reported experiment we now plan a series of analyses aimed at the development of personalized models of emotion.
We are planning to analyze the data from the first phase of the experiment to develop certain calibration and personalization methods to fine tune the interpretation of emotion of in several tasks.
While we are currently experimenting with video games, we are also considering the incorporation of the developed models into decision support and recommendation systems.

\ack The authors are grateful to Academic Computer Centre CYFRONET AGH for granting access to the computing infrastructure built in the projects No. POIG.02.03.00-00-028/08 ``PLATON -- Science Services Platform'' and No. POIG.02.03.00-00-110/13 ``Deploying high-availability, critical services in Metropolitan Area Networks (MAN-HA)''.

\bibliography{geistbib/geisteam,geistbib/geistpub,%
              afcbib/afcaiteam,afcbib/afcaipub}

\begin{thebibliography}{10}

\bibitem{handbookofemotions}
{\em Handbook of Emotions}, eds., Lisa~Feldman Barrett, Michael Lewis, and
  Jeannette~M. Haviland-Jones, The Guilford Press, New York, NY, 4th edn.,
  2016.

\bibitem{bradley2007iads}
M.~M. Bradley and P.~J. Lang, `The international affective digitized sounds
  (2nd edition; iads-2): Affective ratings of sounds and instruction manual.
  technical report {B}-3', Technical report, University of Florida, Gainsville,
  FL, (2007).

\bibitem{calvo2015}
{\em The Oxford Handbook of Affective Computing}, eds., Rafael~A. Calvo,
  Sidney~K. D'Mello, Jonathan Gratch, and Arvid Kappas, Oxford Library of
  Psychology, Oxford University Press, Oxford, 2015.

\bibitem{costa1992neoffi}
P.T. Costa and R.R. McCrae, {\em Revised NEO Personality Inventory (NEO-PI-R)
  and NEO Five Factor Inventory (NEO-FFI). Professional manual}, Psychological
  Assessment Resources, Odessa, FL, 1992.

\bibitem{hook2008affective}
Kristina H{\"{o}}{\"{o}}k, `Affective loop experiences - what are they?', in
  {\em Persuasive Technology, Third International Conference, {PERSUASIVE}
  2008, Oulu, Finland, June 4-6, 2008. Proceedings}, eds., Harri
  Oinas{-}Kukkonen, Per F.~V. Hasle, Marja Harjumaa, Katarina Segerst{\aa}hl,
  and Peter {\O}hrstr{\o}m, volume 5033 of {\em Lecture Notes in Computer
  Science}, pp. 1--12. Springer, (2008).

\bibitem{ijsselsteijn2013geq}
W.A. IJsselsteijn, Y.A.W. {de Kort}, and K.~Poels, {\em The Game Experience
  Questionnaire}, Technische Universiteit Eindhoven, 2013.

\bibitem{james1884what}
William James, `What is an emotion?', {\em Mind}, {\bf 9}(34),  188--205,
  (1884).

\bibitem{johnson2018pensgeq}
Daniel~M. Johnson, M.~John Gardner, and Ryan Perry, `Validation of two game
  experience scales: The player experience of need satisfaction {(PENS)} and
  game experience questionnaire {(GEQ)}', {\em Int. J. Hum. Comput. Stud.},
  {\bf 118},  38--46, (2018).

\bibitem{afcsensors-icaisc2018}
Krzysztof Kutt, Wojciech Binek, Piotr Misiak, Grzegorz~J. Nalepa, and Szymon
  Bobek, `Towards the development of sensor platform for processing
  physiological data from wearable sensors', in {\em Artificial Intelligence
  and Soft Computing - 17th International Conference, {ICAISC} 2018, Zakopane,
  Poland, June 3-7, 2018, Proceedings, Part {II}}, pp. 168--178, (2018).

\bibitem{kkt2019afcai}
Krzysztof Kutt, Dominika Dr\k{a}\.{z}yk, Pawe\l{} Jemio\l{}o, Szymon Bobek,
  Barbara Gi\.{z}ycka, V\'{\i}ctor~Rodr\'{\i}guez Fern\'{a}ndez, and
  Grzegorz~J. Nalepa, `{BIRAFFE}: Bio-reactions and faces for emotion-based
  personalization', in {\em AfCAI 2019: Workshop on Affective Computing and
  Context Awareness in Ambient Intelligence}, volume 2609 of {\em {CEUR}
  Workshop Proceedings}. CEUR-WS.org, (2020).

\bibitem{lang2008iaps}
P.~J. Lang, M.~M. Bradley, and B.~N. Cuthbert, `International affective picture
  system (iaps): Affective ratings of pictures and instruction manual.
  technical report {B}-3', Technical report, The Center for Research in
  Psychophysiology, University of Florida, Gainsville, FL, (2008).

\bibitem{law2018geq}
Effie L.-C. Law, Florian Br\"{u}hlmann, and Elisa~D. Mekler, `Systematic review
  and validation of the game experience questionnaire (geq) - implications for
  citation and reporting practice', in {\em Proceedings of the 2018 Annual
  Symposium on Computer-Human Interaction in Play}, CHI PLAY ’18, p.
  257–270, New York, NY, USA, (2018). Association for Computing Machinery.

\bibitem{gjn2019fgcs}
Grzegorz~J. Nalepa, Krzysztof Kutt, and Szymon Bobek, `Mobile platform for
  affective context-aware systems', {\em Future Generation Computer Systems},
  {\bf 92},  490--503, (mar 2019).

\bibitem{gjn2019sensors}
Grzegorz~J. Nalepa, Krzysztof Kutt, Barbara Gi\.zycka, Pawe\l{} Jemio\l{}o, and
  Szymon Bobek, `Analysis and use of the emotional context with wearable
  devices for games and intelligent assistants', {\em Sensors}, {\bf 19}(11),
  2509, (2019).

\bibitem{peirce2019psychopy}
Jonathan Peirce, Jeremy~R. Gray, Sol Simpson, Michael MacAskill, Richard
  H{\"o}chenberger, Hiroyuki Sogo, Erik Kastman, and Jonas~Kristoffer
  Lindel{\o}v, `Psychopy2: Experiments in behavior made easy', {\em Behavior
  Research Methods}, {\bf 51}(1),  195--203, (2019).

\bibitem{picard1997affective}
Rosalind~W. Picard, {\em Affective Computing}, MIT Press, Cambridge, MA, 1997.

\bibitem{dooren2012gsr16locations}
Marieke van Dooren, J.~J.~G. de~Vries, and Joris~H. Janssen, `Emotional
  sweating across the body: Comparing 16 different skin conductance measurement
  locations', {\em Physiology \& Behavior}, {\bf 106}(2),  298 -- 304, (2012).

\bibitem{heartpy}
Paul van Gent, Jonathan de~Bruin, Kris, and Glenn Fernandes.
\newblock {H}eart{P}y -- {P}ython {H}eart {R}ate analysis toolkit.
\newblock \url{https://doi.org/10.5281/zenodo.1324310}.

\bibitem{ptp1998neoffi}
B.~Zawadzki, J.~Strelau, P.~Szczepaniak, and M.~\'Sliwi\'nska, {\em Inwentarz
  osobowo\'sci NEO-FFI Costy i McCrae. Polska adaptacja}, Pracowania Test\'ow
  Psychologicznych PTP, Warszawa, 1998.

\bibitem{zuchowska2020eng}
Laura \.Zuchowska, {\em Game Design with Unity for Affective Games}, {BS}c
  thesis, AGH University of Science and Technology, 2020.
\newblock Supervisor: G.J. Nalepa.

\end{thebibliography}

\end{document}